\documentstyle[aps,preprint,tighten,floats,epsf,rotate]{revtex}
\normalsize

\def\be{\begin{equation}} 
\def\ee{\end{equation}} 
\def\bea{\begin{eqnarray}} 
\def\eea{\end{eqnarray}} 
\def\gev{{\rm \,Ge\kern-0.125em V}}
\def\mev{{\rm \,Me\kern-0.125em V}}
\def\lsim{\mathrel{\vcenter{\hbox{$<$}\nointerlineskip\hbox{$\sim$}}}}
\def\gsim{\mathrel{\vcenter{\hbox{$>$}\nointerlineskip\hbox{$\sim$}}}}
\def\mpl{M_{Pl}}
\def\r100{r_{100}}
\def\gama100{\gamma_{100}}
\def\rca{r_{a}}
\def\rcb{r_{c}}

\def\v100{v_{100}}

\begin{document}
\bibliographystyle{unsrt}
\footskip 1.0cm
\thispagestyle{empty}
\vspace*{10mm}
\centerline {\Large Electroweak Baryogenesis in a Cold Universe}
\vspace*{8mm}
\centerline {\large Raghavan Rangarajan}
\centerline {\it Theoretical Physics Division} 
\centerline {\it Physical Research Laboratory} 
\centerline {\it Navrangpura, Ahmedabad 380 009, INDIA}
\vspace*{5mm}
\centerline {\large Supratim Sengupta and Ajit M. Srivastava}
\centerline {\it Institute of Physics} 
\centerline {\it Sachivalaya Marg, Bhubaneswar 751005, INDIA}
\vspace*{8mm}
\date{\today} 
\baselineskip=18pt
 
\centerline {\bf ABSTRACT}
\vspace*{4mm}
 
 We discuss the possibility of generating the baryon asymmetry of the
Universe when the temperature of the Universe is much below the
electroweak scale. In our model the evaporation of primordial black
holes or the decay of massive particles re-heats the surrounding plasma
to temperatures above the electroweak transition temperature leading
to the restoration of electroweak symmetry locally. The symmetry is broken
again spontaneously as the plasma cools and a baryon asymmetry
is generated during the phase transition. This mechanism generates 
sufficient asymmetry for a second order electroweak phase transition.  
For a first order phase transition, sufficient asymmetry is generated 
if viscous effects slow down the heated plasma as it moves away
from the black hole.  In our scenario there is no wash-out of the asymmetry 
after the phase transition as the plasma rapidly cools to lower temperatures 
thereby shutting off the sphaleron processes.

\newpage

\section{Introduction}

In recent years much effort has been devoted to formulating a mechanism
for baryogenesis at the electroweak phase transition
\cite{ckn}. Electroweak baryogenesis is indeed a very exciting 
possibility as here one is working within
the framework of a theory (or its extensions) which is reasonably well 
understood and the energy scale involved is accessible in
laboratory experiments. However the requirement of a strong first 
order phase transition, large CP violation 
and a small Higgs mass to ensure that the created baryon
asymmetry is not washed away after the phase transition places 
stringent constraints on these models and practically rules out
electroweak baryogenesis in the context of the Standard Model.
In this paper we discuss an alternative scenario for electroweak 
baryogenesis. Here the baryon asymmetry is created at temperatures 
much below the electroweak transition temperature
during the evaporation of primordial black holes.
When a black hole is evaporating it heats
up the plasma around it to a temperature much higher than the ambient 
temperature for a short time.  This can also happen due to the decay of
massive particles.  For appropriate black hole masses (or, particle masses) 
the temperature of the hot region  rises above the electroweak transition 
temperature $T_{ew}$ and the electroweak symmetry is restored locally.
Due to the transfer of energy out of this region the hot region will 
cool and the temperature will fall below $T_{ew}$.  Thus in these 
hot regions the electroweak phase transition occurs again.  Baryon asymmetry
is then generated in these hot regions during this phase transition.

One motivation for our model is that 
a first order transition at a very early stage in
the Universe, or density fluctuations in general, seem to inevitably lead
to the formation of primordial black holes. Evaporation of these black holes
will lead to local heating of the plasma and hence to the possibility of
electroweak baryogenesis for appropriate black hole masses.
Also, scalars found in string theory can
decay after the electroweak transition and dilute any baryon asymmetry created 
earlier \cite{dilution}.  But in our scenario, we envision that the 
particles that dilute the baryon asymmetry during their decay can also 
recreate the baryon asymmetry after their decay. 

The basic mechanism  of our model can be realized either by particles emitted
in the evaporation of a primordial black hole or by particles constituting
the decay products of a heavy particle. In both cases, it is the energy of  
these emitted particles which thermalizes and leads to local heating
of the surrounding plasma. For the black hole case the heated region
will be spherically symmetric while for the particle decay case the region 
will be collimated. In this paper we will only discuss the black hole case.
The particle decay case is much more complicated. We will briefly comment
on this case at the end of the paper but it is not clear to us
at this stage if in that case one can get sufficient heating of the local 
regions. We hope to discuss the particle decay case in a future work.

A desirable feature of our scenario is that it works %for a first order as well
for a second order transition. In the standard cosmological scenario 
where the Universe cools due to Hubble expansion, a second order phase 
transition can not produce sufficient baryon asymmetry as the baryon 
asymmetry produced is proportional to $\dot \phi/\phi \sim H$, where
$\phi$ is the Higgs field, and this quantity is too small%
\footnote{
However models of electroweak baryogenesis with topological defects are
insensitive to the order of the electroweak phase transition and can create
the observed asymmetry for a second order phase transition \cite{brand}.}.  
In our scenario the heated plasma cools much faster as it moves away from 
the black hole and $\dot \phi/\phi$ is large enough to produce the observed 
baryon asymmetry.  For a first order phase transition, the asymmetry generated
is suppressed by the Lorentz factor associated with the relativistic
speeds at which the heated plasma moves away from the black hole.  However
if viscous effects in the plasma slow down the outward moving plasma
this suppression will be
decreased.  In our scenario, since the heated plasma cools rapidly, the 
temperature after baryogenesis quickly becomes much smaller than the 
electroweak scale preventing any significant wash-out of the created 
asymmetry after the phase transition. 

  We divide the calculation of the baryon asymmetry in our 
model into two main steps. In Section 2, we estimate the 
volume of the plasma in which the electroweak symmetry is restored 
and which is relevant for baryogenesis in our scenario.
In Section 3, we estimate the baryon asymmetry generated as these regions
cool and undergo the electroweak transition. 
Then in Section 4 we point out why the upper bound on the Higgs mass does not
exist in our model.  We conclude in Section 5 with a brief discussion of how
inhomogeneities in the baryon number distribution can be smoothed out before
nucleosynthesis.  

We briefly comment on the formation of primordial black holes in the early 
Universe.  In pre-inflationary days it was argued that the initial spectrum 
of density inhomogeneities, invoked to explain the observed structure in the 
Universe, could also give rise to primordial black holes \cite{hawking,carr}.
It was also argued that white holes would 
be unstable and would convert to black holes \cite{eardley}.
In first order phase transitions, particularly in first order inflationary
scenarios, primordial black holes can be produced by collapsing regions of 
false vacuum trapped between bubbles of the true 
vacuum \cite{kodamaetal,hsu}.  Primordial black holes can also be 
formed via the gravitational instability of inhomogeneities formed during
bubble wall collisions in first order inflation 
\cite{hawkingetal,barrowetal,moss,khlopovetal}, by large amplitude density 
perturbations produced due to fluctuations in the inflaton 
field \cite{carrlidsey}, by shrinking cosmic string loops 
\cite{hawking2,polzem} and by expanding topological defects produced during 
inflation \cite{bellidoetal}.  As our primary motivation is to demonstrate a 
qualitatively different scenario for implementing electroweak
baryogenesis, we will not go in to the specific mechanism which could
give rise to the formation of black holes of required masses. However
we point out that the formation of black holes which evaporate below the 
electroweak scale has been discussed in \cite{bh3}.  

In ref. \cite{nagatani} a scenario which is similar to ours has been proposed.
We comment later on this work.  The presence of a hot plasma surrounding 
primordial black holes has also been discussed in 
refs. \cite{heckler1,heckler2,cline,kapusta}.  However, we point out that in 
these references the
plasma consists of the black hole radiation and particles produced by the 
interaction of the black hole
radiation with itself.  In contrast, in our scenario we consider the interaction
of the Hawking radiation with the ambient plasma surrounding the black hole,
and the subsequent heating up of this plasma by the Hawking radiation.
The possibility of obtaining the baryon asymmetry of the universe in the plasma
around the black hole is mentioned in refs. \cite{heckler1,kapusta}.

\section{Formation of hot regions}

  A black hole evaporates by emitting Hawking radiation with an associated 
temperature

\be
T_{bh} = {M_{Pl}^2 \over 8 \pi M_{bh}}
\ee

\noindent where $M_{bh}$ is the mass of black hole and $M_{Pl} = 1.2
\times 10^{19} GeV$ is the Planck mass. We use natural units with
$\hbar$ = c = 1. The rate of loss of mass by the
evaporating black hole is given by

\be
{dM_{bh} \over dt} = - {\alpha M_{Pl}^4 \over M_{bh}^2}
\ee

Here, $\alpha$ accounts for the scattering of emitted particles by
the curvature and depends on $T_{bh}$. For different values of
$T_{bh}$ values of $\alpha$ have been tabulated in \cite{scat}.
For $T_{bh}$ = 1, 200 and 10$^{15}$ MeV, the corresponding values of  
$\alpha$ are $3.6 \times 10^{-4}$, $2.3 \times 10^{-3}$ and 4.5 $\times 
10^{-3}$ respectively. As we will see later, the relevant value of 
$T_{bh}$ for us will be higher than about 10$^6$ GeV. Therefore we shall 
set $\alpha$ to be $3 \times 10^{-3}$.

The lifetime $\tau_{bh}$ of the black hole can be obtained by 
integrating eqn.(2). We get
\be
\tau_{bh} \simeq 10^2  M_{Pl}^{-4} M_0^3
\ee
\noindent where $M_0$ is the initial mass of the black hole.
Eqn.(2) implies that very little energy is emitted until time of the order of
$\tau_{bh}$ which is when  most of the energy of the black hole gets emitted.
Thus for a black hole formed early in the Universe, it is reasonable to 
assume that the black hole essentially evaporates only when the age of the
Universe is of order $\tau_{bh}$. 

Let us assume that black holes formed
in the early Universe have masses so that they evaporate when the
temperature of the Universe is $T_U<T_{ew}$. 
The age of the Universe $t_U$ when its temperature is $T_U$ is

\be
t_U \simeq 0.3 g_*^{-1/2} M_{Pl} T_U^{-2}
\ee

Here $g_*$ is the number of degrees of freedom relevant at temperature $T_U$. 
We are interested in black holes that decay after the electroweak phase   
transition but before the onset of nucleosynthesis at $T_U=1\mev$.  For
concreteness, we shall consider black holes decaying at $T_U$ equal to
1 \gev and 10 \gev.  For $T_U$ in the range 1-100 GeV, $g_*$ is 
equal \cite{kolb} to 100.  From eqn.(3) we can then calculate $M_0$ so that 
$\tau_{bh} = t_U$. We get (by using $g_*$ = 100),

\be
M_0 =0.07 M_{Pl}^{5/3} T_U^{-2/3}
\label{eq:M0}
\ee

The temperature of this black hole is

\be
T_{bh} = 0.6 M_{Pl}^{1/3} T_U^{2/3}\, .
\ee  
 
For $T_U$ = 1 GeV we get $M_0 = 4 \times 10^{11} M_{Pl}$, $T_{bh}
= 1\times 10^6$ GeV and $\tau_{bh} = 5\times 10^{17}\gev^{-1}=
3 \times 10^{-7}$ s.  The picture then is that these black holes emit 
particles with energies roughly equal to $10^6$ GeV into the background 
plasma which is at a temperature $T_U$ ($\sim$ 1 GeV) to start with. 
These $10^6$ GeV particles will scatter with the particles 
in the background plasma and will heat it up through their energy loss. 
For the black hole masses considered here 
only elementary particles will be emitted,
such as quarks, gluons, photons, leptons, etc. (Emission of quarks
and gluons by black holes is a non-trivial process as discussed in
\cite{qg}. We will not worry about those details here.) 
For $T_U=10\gev$, $M_0=8\times10^{10}\mpl$, $T_{bh} = 6\times 10^6$ GeV 
and $\tau_{bh} = 4\times 10^{15}\gev^{-1}= 3 \times 10^{-9}$ s.

Obtaining the temperature profile outside a black hole radiating into
an ambient plasma is non-trivial.  We first show below how energy from
the black hole is deposited in the surrounding plasma thereby heating it up.
It is necessary to include heat transfer to allow for 
the energy deposited close to the black hole to move out.  We first
use diffusion equations from stellar physics but then raise a concern that
one does not have hydrostatic equilibrium in the plasma surrounding the black
hole, unlike in stellar atmospheres.  Therefore the plasma also moves out due
to the pressure gradient.  We model this outflow of the plasma using shells.
However a more complicated numerical study may be necessary to understand
the simultaneous heating of the plasma and energy transfer due to diffusion 
and bulk flow.

The energy loss of particles traversing a region of quark-gluon plasma has 
been discussed extensively in the literature.
The energy loss per unit distance for an energetic quark with 
energy $E$ traversing a quark-gluon plasma at temperature $T$ is given by%
\footnote{%
This expression was derived
for energies lower than $10^6\gev$ but since $\alpha_s$ is less for higher $E$
the perturbative derivation of the energy loss is even more accurate
for our case.}
\cite{eloss}

\be
-{dE \over dx} \approx 0.1 T^2 ln(\sqrt {E/T}) \, ,
\ee
where we have taken $\alpha_s\sim 0.1$.  For the black hole of mass 
$4\times10^{11}M_{Pl}$, setting $T$ equal to the ambient temperature 
$T_U=1\gev$ and $\Delta E=10^6 \gev$, we get $\Delta x\sim10^6 \gev^{-1}$,
if we approximate $E(x)$ as $T_{bh}$.  One could argue that $\Delta x$ is 
the distance in which the energy released by the black hole is thermalised
as by $\Delta x$ an emitted particle has attained an energy comparable to 
the thermal energy of the particles in the surrounding plasma (loosely 
speaking, it has thermalised).  Furthermore, it takes many collisions before
the radiated particle comes to the same energy as the background, in which 
time the heat transferred along the way would also thermalise due to 
collisions.  If one now equates $M_{bh}={\pi^2\over30} g_* T^4 {4\pi\over3} 
r_{st}^3$ where the stopping distance $r_{st}=\Delta x$ then $T=400\gev$.
However we are here grossly assuming that all the emitted energetic particles 
go through a plasma at 1\gev and that all the black hole energy is
thermalised to a uniform temperature.  Instead let us consider the black hole 
evaporation in steps and equate $0.01 M_{bh}$ with the increase in energy of 
the ambient plasma within $r_{st}$.  The temperature of the plasma rises to 
100\gev.  The next 1\% of the black hole mass released as Hawking radiation 
will now see an ambient plasma at 100\gev and hence the new stopping 
distance $r_{st}^\prime$ will be $200\gev^{-1}$.  Thus a smaller region than 
before is heated to an even higher temperature.  This continual process can 
give rise to a temperature gradient in the plasma surrounding the plasma.
The above also shows that naively equating the energy emitted with
$\rho {4\pi\over3} r_{st}^3$, as we did above, will not be correct.
Though the plasma within $r_{st}$ rises to $T_{ew}$ even with the energy of
1\% of the black hole we do not use this to estimate the volume in which
the symmetry is restored as it is not clear what fraction of the black
hole mass, emitted as radiation, sees a plasma at the initial ambient 
temperature of 1\gev.

To obtain the temperature profile outside the black hole 
one must include heat transfer from the regions close to the black hole to
outer regions.  (In fact, if we ignore heat transfer the temperature of
the plasma close to the black hole keeps rising until the plasma temperature
becomes equal to the black hole temperature.  After this the black hole
can not evaporate any further until the energy deposited close to the
black hole moves out.) To include the effects of diffusion we shall
follow the calculation of the temperature profile of stellar interiors.
In stars one has a hot stellar core in which nuclear reactions occur,
and a stellar atmosphere surrounding the core.  Energy moves through the 
stellar atmosphere via diffusion due to a thermal gradient.
We have an analogous situation with a black hole radiating into a 
plasma around it  and we shall assume that energy radiated by the black hole 
is quickly thermalised, so at any given time one has a hot black hole 
radiating into a plasma in which there is a thermal gradient.  Energy 
diffuses from the hotter regions closer to the black hole to regions further 
away due to the thermal gradient.  Under these assumptions we will use eqn.
5.11 of \cite{KW} (hereafter referred to as KW), namely,
\be
\partial T/\partial r=-{3\over 16\pi a c} {\kappa \rho l\over r^2 T^3}
\, ,
\label{eq:dTdr}\ee
where $c=1$ is the speed of light, $a=(\pi^2/30) g_*$.  $\kappa\rho$ is the 
radiative cross-section per unit mass averaged over frequency times the  
energy density and equals $\sigma n=$1/mean free path,
where $\sigma$ is the cross section for scattering and
$n$ is the number density of particles.  The local luminosity
$l(r)$ is defined as the net energy passing outward
through a sphere of radius $r$ per second (see Section 4.2 of KW).

To integrate $\partial T/\partial r$, we need $\sigma n$ and $l(r)$.
We use 
\be
\sigma n= (1/9)(2) {4 \pi \alpha_s^2\over 3(4T^2)}\times (3/4)(1.2/\pi^2)g_*T^3
\, .  
\ee
The above expression for $\sigma$ is the cross section for 
quark-antiquark scattering (see eqn. 6.33 of \cite{HM}). 
The factors of (1/9)(2) come from averaging over color and
tracing over QCD generators, respectively.  We have set the 
Mandelstam variable $s$ in the formula to equal $(2T)^2$, i.e., for incoming
particles of equal (and opposite) momenta $T$.  Then we find that
$\sigma n=2\times10^{-4}g_* T=0.02 T,$ where we have used $\alpha_s=0.1$.

In stars $l(r)$ can be a complicated function, depending on the 
distribution of sources and sinks of energy.  Since we are interested
in the steady state situation we shall take the luminosity to be independent
of $r$ and equal to the energy radiated per unit time by the black hole.
Therefore
\be
l(r)=-dM_{bh}/dt 
=\alpha M_{Pl}^4/M_{bh}^2\,.  
\ee
Now one can integrate $\partial T/\partial r$, with the boundary condition 
that $T=T^\prime$ at $r=r'$. We shall initially take $T^\prime$ to be
the ambient temperature $T_U$.  A priori one does not know $r'$.
Due to scattering of radiation in the plasma $r'$ should be less than
$c\tau_{bh}$.  We shall fix $r'$ by requiring that the increase in the 
energy of the plasma for $r\le r'$ is equal to the radiated mass of the 
black hole.  We get
\be
T^3-T^{\prime3}=
(9/16\pi a c) {\alpha M_{Pl}^4\over M_{bh}^2} (0.02)
\Biggl[{1\over  r}
- {1\over  r'}\Biggr].
\label{eq:Tprof}\ee
We shall apply this solution for $r\ge6R$, where $R=2M_{bh}/\mpl^2$ is
the Schwarzschild radius of the black hole.  $6R$ is the radius of the thermal
atmosphere around the black hole \cite{zurek} beyond which curvature effects
of the black hole are negligible.

Our assumption of $l(r)$ independent of $r$ implies a steady state profile.
However our choice of the boundary condition that the increase in
energy of the plasma within $r^\prime$ equals the energy radiated by the 
black hole implies that all the black hole energy is used to create the
temperature profile.  More likely, the energy radiated from the black hole
will first create a profile and then the remaining energy of the black hole
will be transmitted through the plasma without further heating the plasma.
This implies that the increase in energy of the plasma will be less than the 
radiated mass of the black hole.  Therefore $T(r)$ for any $r$ will be less 
than what we obtain and our estimates of $r_{100}$ in the diffusion picture 
will be an upper bound.

We start with a black hole of mass $M_{bh}=4\times10^{11}\mpl$
and lifetime $5\times10^{17}\gev^{-1}$ which decays when the temperature
of the universe is 1 \gev.  Using the temperature profile in 
eqn.(\ref{eq:Tprof}) we apply the energy constraint mentioned above to 
(numerically) obtain $r'=2\times10^9\gev^{-1}$,
and thus an expression for $T(r)$.  While applying the energy constraint we
assume that only 90\% of the black hole mass is radiated away.
Using $T(r)$ we then obtain $\r100$, the radius below which the temperature is
greater than $100\gev$, to be $4\times10^2\gev^{-1}$.

Now we envision that the black hole 
has reduced in mass by 90\% and so we have a black hole
of mass $4\times10^{10}\mpl$ and lifetime $\tau=5\times10^{14}\gev^{-1}$.  
We once again use the above equation and find that the energy constraint 
for a black hole that radiates 90\% of its mass is consistent with 
$r'=1\times10^8\gev^{-1}$ and $T(r^\prime) =2\gev$.
(We vary $r'$ and $T(r')$ so that we satisfy the energy condition for
a value of $r'$ such that $T(r')$ is consistent with the temperature profile 
for $M_{bh}=4\times10^{11}\mpl$.  Also, the energy in the region
$r\le r'$ obtained from the temperature profile of the black hole
at $M_{bh}=4\times10^{11}\mpl$ is much smaller than the energy deposited
in this region by the black hole after its mass has reduced to 
$4\times10^{10}\mpl$.) We now find that $\r100=3\times10^4\gev^{-1}$.

Now let the mass of the black hole be $4\times10^9\mpl$. Its lifetime is
$5\times10^{11}\gev^{-1}$. We repeat the procedure above and obtain 
a consistent temperature profile with $T'=30\gev$ at 
$r'=1\times10^6\gev^{-1}$.  Again the energy deposited greatly exceeds the 
energy already in the plasma for $r\le r'$.  We find that 
$\r100=8\times10^5\gev^{-1}$. Since $\r100\sim r^\prime$ we can not get a 
larger $\r100$ as the black hole mass continues to reduce further.
Thus for a black hole of mass $4\times 10^{11}\mpl$ radiating into a plasma
(initially) at a temperature of $1\gev$ we obtain $\r100=8\times10^5\gev^{-1}$.

Had we started with an initial ambient temperature of 10\gev, and a black
hole of mass $8\times10^{10}\mpl$, temperature $6\times10^6\gev$ and lifetime 
$4\times10^{15}\gev^{-1}$ or $3\times10^{-9}$s, we would first
obtain $r'=8\times10^7\gev^{-1}$ for $T'=10\gev$ and 
$r_{100}=8\times10^3\gev^{-1}$.  Allowing the black hole to further radiate 
with a mass of $8\times10^9\gev$ we find $r'=5\times10^6\gev$ with 
$T'=10\gev$.  Then we obtain $\r100$ of $6\times10^5\gev^{-1}$.

If one calculates the energy radiated
per unit time into the black hole by the plasma at $r=6R$, $P=
\sigma (T(6R))^4 4\pi (6R)^2$, where $\sigma=(1/4)(\pi^2/30) g_*$ is the
Stefan-Boltzmann constant modified for more than the two photonic
degrees of freedom, one finds that this is always a factor
of $10^3$ less than $-dM_{bh}/dt|_{vacuum}=\alpha \mpl^4/M_{bh}^2$.  
Therefore we can ignore any decrease in the net luminosity due to
absorption by the black hole.

Though we have used the stellar diffusion equations to obtain the
temperature profile there is one important difference between our situation
and that of stellar atmospheres, namely, the existence of hydrostatic
equilibrium.  While gravity in stars provides hydrostatic equilibrium we 
show below that pressure gradients in the plasma will overcome the 
gravitational attraction of the black hole and the plasma will move out.  
This is to be expected as these black holes have much less mass and plasma 
temperatures are much larger as compared to stellar situations.

The gravitational force (per unit volume) on an element of the plasma of 
density $\rho$ and at a distance $r$ from the black hole is $(\rho +P) g$, 
where $g=G_N M/r^2$ is the gravitational acceleration and $P={1\over3}\rho$ 
is the pressure at a distance $r$ from the black hole.  
$M$ here includes the black hole mass and the 
energy plus pressure of the plasma within the sphere at radius $r$.  
One can show that the contribution of the plasma to $M$ is negligible.  
Therefore the gravitational force (per unit volume) is
\be
f_g={4\pi^2\over 90} g_* {T^4 M_{bh}\over \mpl^2 r^2}\, .
\ee
The outward force (per unit volume)
on the plasma element is given by $\nabla P=\partial P/\partial r$.  Now 
$P={\pi^2\over90}g_* T^4$.  Therefore
\be
\partial P/\partial r={4\pi^2\over90}g_*T^3\partial T/\partial r\,.
\ee
For $M_{bh}=4\times10^{11}\mpl$ we find that outward force due to the pressure
gradient is much larger than the gravitational attractive force
for regions where $T<1\times10^5\gev$.  At $r_{100}$ $\partial P/\partial r$ 
is 9 orders of magnitude larger than the gravitational force.  For 
$M_{bh}=8\times10^{10}\mpl$ gravity becomes sub-dominant at 
$T=7\times 10^5\gev$ and the disparity between the pressure gradient and the 
gravitational force at $\r100$ is even larger. Thus one should include effects 
of bulk flow in the modeling of the plasma outside the black hole.

We now calculate the distance at which heat transfer becomes dominated by
bulk flow rather than diffusion. As we have shown, beyond some 
distance $\rca$ hydrostatic equilibrium is not maintained 
and the plasma acquires some outward bulk velocity.  However bulk plasma motion
will not be the dominant mode of heat transfer till the plasma velocity
increases sufficiently.  Let $\rcb$ be the distance at which
heat transfer due to bulk flow becomes dominant.

Starting from the Euler equation for a relativistic fluid \cite{weinberg}
we obtain the following equation for steady-state situations
(i.e. after setting time derivatives to zero) 
\be
  v \gamma^2 {dv/dr} = - 1/(\rho+P) {\partial P/\partial r}\, ,
\label{eq:dvdr}\ee
where $\gamma = 1/\sqrt{1 - v^2}$ and $c=1$.
Using the expression for the pressure of a relativistic fluid, 
$P = (1/3)\rho$, we get
\be
  v/(1-v^2) dv/dr = -(1/T)(\partial T/\partial r)\, .
\label{eq:dvdt}
\ee
We would like to integrate the above equation with the range of integration 
for $v$ and $r$ being $0$ to $v_{c}$ and $\rca$ to $\rcb$ respectively. 
Since heat transfer in this region is still dominated by diffusion we use
$\partial T/\partial r=-B/(r^2 T^2)$ from eq.~(\ref{eq:dTdr})
above, where $B = (90/(1600\pi^3))0.06 \times 10^{-3}(M_{Pl}^4/M_{bh}^2)$.
We obtain the expression for the temperature profile $T(r)$ as
\be
     T^3 = T_0^3 + 3B(1/r - 1/r_0)\, ,
\label{eq:tempp}\ee
where $T_0=\eta T_{bh}$ and $r_0=6R$.  Since we are trying to ascertain the 
distance $r_{c}$ at which the diffusion approximation breaks down, we can 
not impose a boundary condition at large $r$, i.e., at $r^\prime$ as we have 
done earlier.  However, our ignorance of the plasma temperature 
at small $r=6R$ leads us to introduce the parameter $\eta$. 
$\eta$ is then a measure of how different the temperature at $r=6R$ is from 
the black hole temperature $T_{bh}$.  To know the precise value of $\eta$ would
require a complicated numerical simulation of the processes occurring close to
the black hole.  Therefore we instead solve for $\rcb$ for different values of
$\eta$.  For a black hole of mass $4\times 10^{11}\mpl$, if we substitute 
values of $\eta \ge 0.12$ in eq.~(\ref{eq:tempp}) then even at
$r=\infty$ the temperature is greater than $T_{a}=1\times 10^5\gev$.  Since
hydrostatic equilibrium breaks down for temperatures below this value this
implies that hydrostatic equilibrium is maintained for all $r$ and that the
diffusion approximation is valid everywhere.  However the large value of 
$T$ at infinity for these values of $\eta$ implies that
the entire universe is heated up by a single black hole which violates
energy conservation and shows that $\eta \ge 0.12$ should not be used.
For $\eta=0.10$, $\rca$ is $6R=4\times10^{-7}\gev^{-1}$ and so for smaller
values of $\eta$ one also takes $\rca$ equal to $6R$.  For values of
$\eta$ between 0.10 and 0.12 $\rca$ is greater than $6R$ and is obtained
from eq.~(\ref{eq:tempp}) to be 
$3.0\times10^{-10}(1.8\times10^{-3}-\eta^3)^{-1}\gev^{-1}$.
For a black hole of mass $8\times10^{10}\gev$ values of $\eta\ge0.13$ violate
energy conservation while for $\eta\le0.12$ we take 
$\rca$ as $6R=8\times10^{-8} \gev^{-1}$.  For values of
$\eta$ between 0.12 and 0.13 $\rca$ is $3.4\times10^{-11}(2.0\times10^{-3}
-\eta^3)^{-1}\gev^{-1}$.
 
$v_{c}$ is then obtained from eq.~(\ref{eq:dvdt}) as
\be
 v_{c}^2 = 1 - exp\biggl(-\int_{\rca}^{\rcb}2B/(r^2 T^3) dr\biggr)\, .
\label{eq:vcb}\ee

$r_{c}$ can be obtained by equating the flux due to bulk flow at $r_{c}$,
$J(\rcb) = \gamma^2(\rho+P)(\rcb) v_{c}$, with the total flux at $r_{c}$, i.e.,
\be     
           \gamma^2 {2\pi^2\over45} g_* T_{c}^4 v_{c} = l/(4\pi r_{c}^2)
\label{eq:fluxeq}
\ee
where $T_{c}$ is the temperature at $r_{c}$ and $l$ is the luminosity at 
$r_{c}$, which we take to be $-dM_{bh}/dt$.  This finally gives 
\be
\gamma^2 {2\pi^2\over45} g_* T_{c}^4 r_{c}^2 v_{c} = 
(\alpha M_{Pl}^4)/(4\pi M_{bh}^2)
\label{eq:rc}
\ee
By substituting the expressions for $v_{c}$ and $T_{c}$ 
obtained in terms of $r_{c}$, it is
in principle possible to obtain $r_{c}$. 
However the above equation cannot be solved analytically. Instead we 
numerically estimate $r_{c}$ by plotting (LHS-RHS) of eqn.(\ref{eq:rc}) as a 
function of $r_{c}$ to ascertain at what point the graph crosses the 
$r_{c}$-axis.  Substituting this value into eqs.~(\ref{eq:vcb}) and 
(\ref{eq:tempp}) gives $v_{c}$ and $T_c$ respectively.

As mentioned earlier, we present results for certain values of $\eta$.
For a black hole of mass $M_{bh} = 4\times10^{11} M_{Pl}$ evaporating when the 
ambient temperature of the Universe is $T_{U}=1\gev$ we find 
$r_c =9\times10^{-6}\gev^{-1}$ and $3\times10^{-6}\gev^{-1}$ for 
$\eta=0.10$ and $0.08$ respectively.  The corresponding values of $v_c$ are 
0.5 and 0.8 respectively.  The corresponding values of $T_c$ are
$1\times10^5\gev$ and $6\times10^4\gev$ respectively.
For a black hole of mass $M_{bh} = 8\times 
10^{10} M_{Pl}$ evaporating when the ambient temperature 
of the Universe is $T_{U}=10 \gev$ we find
$r_c=2\times10^{-6}\gev^{-1}$ and $5\times10^{-6}$ for $\eta=0.10$ and $0.08$
respectively.  The corresponding values of $v_c$ are 0.5 and 0.8 respectively.
The corresponding values of $T_c$ are
$5\times10^5\gev$ and $3\times10^5\gev$ respectively.
In all the above cases the temperature of the plasma at $r_c$ is much greater
than 100\gev indicating that $r_c$ is smaller than $\r100$. 
This shows that bulk flow dominates over diffusive transfer of energy in 
the plasma before the temperature of the plasma drops down to the 
electroweak transition temperature, thereby calling in question the validity
of the diffusion picture.

To study the outward motion of the plasma for $r>r_{c}$ we shall consider 
the plasma as consisting of outgoing shells of infinitesimal thickness 
$dr$ (in the frame of the black hole).  As these shells
move out, they cool down due to the expansion of the plasma. Eventually
these shells will reach a distance $\r100$ where the temperature (as
defined in the frame of the plasma) becomes
smaller than 100\gev. At this time the electroweak transition will
occur in these shells and, as we will discuss later, baryogenesis can take 
place there.  
Rewriting eqn.(\ref{eq:dvdt}) and integrating we get
\be
     v^2 = 1 - (T/T_{c})^{2}(1-v_{c}^2)\, , 
\ee
which gives us a relation between the velocity of the outward moving plasma
shell and the temperature of the plasma.

Since the temperature decreases with increasing $r$, 
the velocity of the plasma shell will keep increasing and will eventually
become very close to the speed of light.  The Lorentz factor 
is given by
\be
\gamma={1\over\sqrt{1-v_c^2}}{T_c\over T}\,.
\label{eq:lorentz}
\ee
For $T_U=1\gev$ the Lorentz factor $\gama100$ at $\r100$ is
1000 for $\eta=0.10$ and $0.08$.  
For $T_U=10\gev$, $\gama100$
is 6000 and 5000 for $\eta=0.10$ and $0.08$ respectively.
Below we shall take $v_{100}$, the velocity of the plasma shell at $\r100$,
to be 1.  Later we shall briefly mention
the possible role of viscosity in slowing down the plasma and its consequences.

Assuming a steady state situation $l(r)$ is independent
of $r$ and equal to $-dM_{bh}/dt$.  Therefore 
for $r\ge r_c$ 
\be
r=\Biggl[{l\over \gamma^2 (8\pi^3 g_*/45) T^4 v }\Biggr]^{1\over2} \,.
\label{eq:lumin}\ee
Above 
$T$ is the 
temperature of the plasma shell at a distance $r$.
For a black hole of mass $4\times 10^{11}\mpl$,
$\r100$ is $7\gamma^{-1}\gev^{-1}$ and it is
$30\gamma^{-1}\gev^{-1}$ for a black hole of mass $8\times 10^{10}\mpl$.

The volume relevant for baryogenesis must be calculated in the rest frame of
the plasma since the sphaleron rates have been calculated in this frame. 
Consider a plasma shell of radius $r_{100}$ in the black hole
frame.  There is no frame in which the
entire shell is at rest.  Instead one can divide the shell into plasma elements
of area $d\Omega r_{100}^2$ and thickness $dr$.  In the rest frame of such a
plasma element the volume element is $d\Omega r_{100}^2 dr'$, where 
$dr'=\gamma dr$ is the shell thickness in the frame of the plasma element.  
Therefore
the volume relevant for baryogenesis contributed by one shell is
\be
dV= 4\pi r_{100}^2 dr' = 4 \pi r_{100}^2  \gama100 dr
= 4 \pi r_{100}^2  \gama100 \v100 dt \,.
\ee
Plasma shells keep moving across $\r100$ for a duration of the order of the 
lifetime $\tau$ of the black hole.  Thus
the total volume relevant for baryogenesis is then given by
\be
V = 4\pi r_{100}^2 \gama100 \v100 \tau
=3 \gama100^{-1} l \tau/(4 \rho_{100}) \, ,
\ee
where we have used eq.~(\ref{eq:lumin}) for the second equality.
Since $l=-dM_{bh}/dt=\alpha M_{Pl}^4/M_{bh}^2$ we may also express the volume as
\be
V=7\times 10^{-11} \gev^{-4} \gama100^{-1}M_{bh}
 =5\times10^{-12} \gev^{-4}\gama100^{-1} M_{Pl}^{5/3} /T_U^{2/3}
\ee
where we have used eq.~(\ref{eq:M0}) for the last equality.
For a black hole of mass $4\times 10^{11}\mpl$ radiating into a plasma
initially at $1\gev$, $V$ is $3\times10^{20}\gama100^{-1}\gev^{-3}$; 
for a black hole of mass $8\times 10^{10} M_{Pl}$ and an
initial ambient temperature of 10\gev, 
$V$ is $7\times10^{19}\gama100^{-1}\gev^{-3}$.

Electroweak baryogenesis takes place via sphaleron processes with
sphaleron size $\sim (\alpha_W T_{ew})^{-1} \simeq$ (3 GeV)$^{-1}$. %Thus
Since the thickness of the shell in the plasma frame for a temperature
difference of 1\gev,i.e., $r_{99}'-r_{100}'$, is proportional to a
Lorentz factor, which compensates for the $\gamma$ factor in $r_{100}$,
one can easily fit the sphaleron in the plasma shell 
at $r_{100}$.

\section{Estimation of the baryon asymmetry}

Let us assume that the number of excess baryons created around each black hole
is $N_B$. If the number density of black holes is $n_{bh}$
then the number density $n_B$ of baryon excess produced by these
black holes is $n_B = N_B n_{bh}$. To estimate $n_{bh}$ we 
assume that the net energy density of black holes does not dominate
the energy density of the Universe at the time when black holes
decay.  (With this assumption we are assured that black holes never 
dominate the dynamics of expansion of the Universe so the overall picture 
of the evolution of the Universe remains unchanged.) 
This gives the following expression for $n_{bh}$ (using eqn.(5)).
\be
n_{bh} \lsim  5 \times 10^2 T_U^{14/3} M_{pl}^{-5/3}
\ee
For $T_U= 1\gev$, $n_{bh}\lsim  8\times 10^{-30}\gev^3$, while for
$T_U= 10\gev$, $n_{bh}\lsim  4\times 10^{-25}\gev^3$.  We shall use
the maximum allowed value of $n_{bh}$ below.

The entropy density of the Universe (which will remain essentially
unchanged even after all the black holes evaporate, since the black holes
do not dominate the energy density of the Universe) is given by

\be
s = {2 \pi^2\over 45} g_* T_U^3 \simeq 40 T_U^3 
\ee

We now proceed to estimate the baryon asymmetry produced in the
hot regions by models of baryogenesis conventionally used at the electroweak 
scale.  In the following we shall use V to refer to the hot region where the
temperature rises to at least $T_{ew}$ as well as to the volume of this region. 
Let us first assume that the electroweak transition is of first order. 
Before the evaporation of the black hole the Universe is in a state with the 
expectation value of $\phi$ at a value $v$.  When the regions of volume V 
heat up, the expectation value of $\phi$ in V will evolve to the value 0.
(Later we consider the possibility that $\phi$ may not roll to the minimum
due to insufficient heating.) As the plasma in V cools further, a local 
minimum at a non-zero value of $\phi$ appears in the effective potential.
For $T<T_{ew}$, this minimum becomes the true minimum of the potential.  
Bubbles of true vacuum now appear in the hot region V as the field tunnels 
from 0 to $v$.  As these bubbles grow, they either produce a
baryon asymmetry in the outer regions of a thick wall by ``spontaneous 
baryogenesis'' \cite{spontbary} or by the asymmetric reflection 
of particles in the symmetric phase from the surface 
of a thin bubble wall \cite{parttrans}.  In the former case, the non-zero value
of the time derivative of a field coupled to the baryon number acts as a 
chemical potential for baryon number, thereby biasing B+L violating
sphaleron processes in the wall to produce more baryons than anti-baryons.
In the latter case, the asymmetry created in some quantum number not 
orthogonal to baryon number due to reflections off the bubble wall is 
converted into a baryon asymmetry by sphaleron processes.  This baryon 
asymmetry is frozen after it passes through the wall.  A more sophisticated 
treatment of CP violation, particle transport and diffusion
in the context of electroweak baryogenesis is discussed in
refs. \cite{ckndiff,comellietal,huetnelson1,huetnelson2,prokopecetal}. 
A baryon asymmetry can also be produced at the electroweak phase transition
through the CP asymmetric interaction of the bubble wall with fluctuations
in the winding number of gauge-Higgs fields configurations \cite{TZ,MSTV}
and through the interference between electroweak sphaleron-induced baryon
number violating processes and QCD sphaleron-induced CP violating processes 
\cite{mclerran}.  In the standard electroweak baryogenesis scenario 
baryogenesis occurs in the entire Universe as the electroweak bubbles sweep 
through the Universe during the phase transition.  In our case, 
our hot regions cover a volume fraction of 
$V n_{bh}= 3 \times 10^{-9}\gamma^{-1} 
\gev^{-4} T_U^4$.  However, the entropy density
of the Universe at a temperature of $T_U$ is also lower by a factor
of $10^6\gev^3/T_U^3$ as compared to the entropy density at the 
electroweak phase transition.  The ratio of the asymmetry that we obtain to 
that in the standard scenario is then volume fraction $\times$ entropy 
factor $=3\times10^{-3} \gamma^{-1}T_U \gev^{-1}$.

The Standard Model does not have sufficient CP violation to create the 
observed baryon asymmetry of the Universe.  Therefore we consider the results 
of ref.\cite{huetnelson2} for baryogenesis in supersymmetric models.  
With CP violating phases of the order of $10^{-4}$ they produce enough 
asymmetry for a first order electroweak phase transition.  
However, even with a larger value of the CP violating parameters 
the asymmetry that
we obtain above for a first order electroweak phase transition will be
insufficient because of the large value of $\gamma$.  

However above we have completely ignored the effects of viscosity.
Viscous effects can be of two types-- bulk and shear\cite{landaulifshitz}.  
The bulk viscosity 
coefficient for a (quark-gluon) relativistic plasma is approximately
0\cite{hosoyakajantie}.  
However there is a contribution to the shear viscosity that is
proportional to $\partial v/\partial r$, which will slow down the plasma
as it moves out.  The relativistic hydrodynamic equations with viscosity
are very complicated and can not be solved easily.  If, however, the viscous
effects decrease the plasma velocity at $\r100$ to even $0.9c$, this will imply
that $\gama100$ is 2.  Then, with a value of CP violation 
$\sim 0.1$ or 0.01,
one can obtain sufficient asymmetry in our scenario for
black holes radiating into an ambient plasma at an initial temperature
of $1\gev$ or $10\gev$ respectively.                                   

Since our formalism above is not very rigorous we shall also estimate below 
the asymmetry in the diffusion picture ignoring any plasma motion due to
pressure gradients.  For a black hole of initial mass $4\times10^{11}\mpl$
the volume fraction $Vn_{bh}$, where $V={4\over3} \pi\r100^3$ is 
$2\times10^{-11}$.  Taking into account the entropy factor, the asymmetry is 
$2\times10^{-5}$ of the asymmetry in standard electroweak baryogenesis models.  
For a black hole of initial mass $8\times10^{10}\mpl$
the volume fraction is $4\times10^{-7}$.
So the asymmetry is about $4\times10^{-4}$ of
the asymmetry in standard electroweak baryogenesis models.  

In the diffusion picture we have ignored the fact that once the black
hole has evaporated completely energy from hotter regions will
diffuse out and heat up cooler regions.  This will effectively increase
$\r100$.  Furthermore, if we relax the constraint that the black holes 
do not dominate the Universe until they decay this may enhance the final 
asymmetry.

We point out an interesting phenomenon that can occur 
in the regions surrounding the black hole where the 
temperature rises but does not get much higher than $T_{ew}$.  In these 
regions the effective potential changes from the zero temperature 
potential to a high temperature potential with a local minimum  at 
$<\phi>=v$ as shown in fig. 1.  In these regions the Universe is trapped in 
a state with $<\phi>=v$ and tunnels to the $<\phi>=0$ vacuum by bubble 
nucleation.  Thus in these regions bubbles of $<\phi>=0$ are created.  
As long as the $<\phi>=v$ vacuum is higher than the $<\phi>=0$ vacuum these 
bubbles expand, as in fig. 2a.  However as this region cools the two vacua 
become degenerate and then the state with $<\phi>=v$ becomes the true vacuum.  
If the $<\phi>=0$ bubbles have not already
disappeared after collisions (leaving the
Universe in the $<\phi>=0$ vacuum state) then the 
$<\phi>=0$ bubbles now begin to shrink, as in fig. 2b.  
Baryons are now created in the inner regions of the shrinking walls.
Furthermore, $<\phi>=v$ bubbles may nucleate inside the $<\phi>=0$ bubbles
(depending on the critical bubble size). 
As these expand baryons are created in their walls. 
The above phenomenon may occur in the diffusion picture for $r>\r100$
in the transition phase after the black hole evaporates away and energy
from hotter regions moves out and heats up colder regions.
In the case of heat transfer through bulk motion of the plasma
this mechanism may occur at some distances away from the black hole
in the early stages before the steady state is achieved. 

We now consider a second order phase transition.
In the hot regions, the symmetry is restored and the field 
settles at the minimum of the potential at $\phi=0$.
As the region cools below $T_{ew}$ the minimum shifts from $\phi = 0$.
$<\dot\phi>$, in this case, is not due to Hubble expansion but is due to the 
cooling of the plasma as the plasma expands.
Hence, unlike in the standard cosmology, 
$<\dot\phi>$ is large.  The non-zero $<\dot\phi>$ acts as a chemical potential 
for baryon number during the phase transition and a net baryon number is 
created by B+L violating sphaleron processes. 
The baryon asymmetry generated in a second order electroweak
phase transition is given by \cite{prokopec}
\be
n_B/s\approx 10^{-19}\epsilon [-\dot T/T]_{100}/H_{100}
\ee
where $H_{100}$ is the value of the Hubble constant when the temperature
of the Universe is 100\gev, $\dot T/T$ is %evaluated at $r_{100}$ 
in the
frame of the plasma when the temperature has fallen to 100 \gev
and $\epsilon$ is a measure of CP violation.

To estimate $\dot T/T$ in the rest frame of the plasma when the temperature
has fallen to 100 \gev %at $r_{100}$ 
we note that
at $r_{100}$ in time $dt$ a shell of thickness $dr=\v100 dt$ 
moves out at a temperature $T_{100}$. As the shell moves out 
$\dot T=
{\partial T\over\partial r}{dr\over dt} {dt\over dt'}
\approx -
Tv\gamma/r$, since eqs.~(\ref{eq:lumin}) and (\ref{eq:lorentz}) imply that 
$T$ is approximately proportional to $r^{-1}$. 
$dt'$ is the time
differential in the plasma frame.
Therefore $[-\dot T/T]_{100}=v_{100} \gama100/r_{100}$.

Using values of $\r100$ and $\v100$ obtained earlier we get
$[-\dot T/T]_{100}=1\times10^{-1} \gama100^2\gev$ and 
$3\times10^{-2}\gama100^2\gev$
for $T_U$
equal to 1\gev and 10\gev respectively.  
In the standard scenario 
$[-\dot T/T]_{100}=H_{100}=10^{-14}\gev$.
Therefore the baryon asymmetry that we obtain in this scenario,
taking into account the volume fraction and the entropy factor, is
\be
n_B/s=3\times10^{-9} \gama100\epsilon \gev^{-1}
\ee
for $T_U=1\gev$ and is 
\be
n_B/s=9\times10^{-9} \gama100\epsilon \gev^{-1}
\ee
for $T_U=10\gev$. Since $\gama100\sim 10^{3}$ for our choices of $\eta$
clearly one can obtain a sufficiently large baryon asymmetry in our
scenario with a second order phase transition.  It is interesting
that while the asymmetry generated in a first order electroweak phase
transition is suppressed by the Lorentz factor,the asymmetry is enhanced 
by the Lorentz factor for a second order phase transition.
Also note that even if viscous effects reduce the value of $\gama100$ 
to 1 we can obtain sufficient asymmetry for $\epsilon\sim 10^{-1}-10^{-2}$
(for $v_{100}$ not much smaller than 1).Further note that the possibility 
of creating a sufficient asymmetry is not very sensitive to the value of 
$\eta$ as long as $T_c$ is large enough ($\gsim 200\gev$) to ensure that 
$v_{100}$ is not much smaller than 1.

Since the distance between black holes is much greater than $r_U$,
the distance at which the outward moving plasma attains the ambient
temperature, the heating of the plasma by one black hole does not directly
affect the plasma around neighboring black holes.  Once the outward
moving plasma reaches $r_U$ it merges with the ambient plasma.  As the
energy density of the black holes does not dominate the energy density 
of the universe the heating of the ambient plasma between black holes, 
which could affect the temperature profile of neighboring black holes, is 
also negligible.  Finally, baryons carried out by the outward moving 
plasma into the ambient plasma can subsequently enter the hot region 
around another black hole due to random Brownian motion across the 
intermediate region.  However as the baryons cross $r_U$ of another black 
hole the plasma moving out from the second black hole would push them out 
thereby preventing them from reaching $r_{100}$ of the second black hole, 
where sphaleron processes in equilibrium could destroy them.

We now briefly comment on the possibility of creating hot regions
by the decay of massive particles. 
Massive particle decays will lead to the development of
a collimated jet of particles, with each particle in the jet contributing
to the heating of the plasma.  From eqn.(7) for the energy loss 
of a particle created by the decay of this massive particle
it seems very difficult to get sufficient energy to heat
the region to electroweak scale. However, note that the expression
for the energy loss as given in eqn.(7) has been derived by neglecting 
very hard collisions. Inclusion of hard collisions may lead to a rapid
branching of the initial decay product which may significantly increase
the energy loss.  Also note that baryogenesis will only take place in regions
that are at least as wide as the sphaleron size.  

\section{Constraints on the Higgs mass}

In the electroweak baryogenesis scenario in the context of the Standard Model 
if the high temperature vev in the broken phase is small then the sphaleron
processes are still active.  These sphaleron processes will destroy the 
baryon asymmetry created during the phase transition.  A large enough high 
temperature Higgs vev in the broken phase to avoid the wash-out of the 
asymmetry implies a small Higgs mass which is tightly constrained by the 
experimental lower bound on the Higgs mass. In our scenario, on the other 
hand, the requirement of a small Higgs mass does not apply, as we show 
below. Of course, this is relevant only if 
viscous forces are sufficient to decrease the suppression of the asymmetry 
due to the Lorentz factor for a first order electroweak phase transition.
A large Higgs mass also implies that the electroweak phase transition is 
likely to be of second order in which case the asymmetry generated at the 
electroweak phase transition in standard electroweak baryogenesis is very 
small \cite{prokopec}.  As we have already indicated, we generate sufficient 
asymmetry with a second order phase transition in our scenario.

The baryon number density after a first order
electroweak phase transition is given by
\be
n_B=n_{Bi} exp(-\int \Gamma dt)
   =n_{Bi} exp(-\int_{100}^{T^\prime} \Gamma/\dot T dT)
\ee
where $n_{Bi}$ is the baryon number density created during the phase 
transition, $\Gamma(T)$ is the rate per unit time of B violation
and we choose the upper limit $T^\prime$ to be $T_U$, the ambient 
temperature of the universe.  More precisely, one
could put $T^\prime$ as the temperature at which the B violation rate becomes
less than the Hubble expansion rate but the above choice is sufficient.
For $\dot T$ we use $-(\gamma v/r)T$ and $r=\r100 T_{100}/T$.  $\gamma$
is given by eqn.(\ref{eq:lorentz}) and we let $v=1$. 

The perturbative rate of baryon number violation in the broken phase is given
by \cite{amc}
\be
\Gamma=y f  \alpha_W^4 T \Biggl[{M_W^7 \over 
\alpha_W^7 T^7}\Biggr] exp[-E_{sph}/T]
\ee
where $E_{sph}= 2 M_W(T)/\alpha_W$, with 
$M_W(T) = M_W(0) \sqrt{1 - T^2/T_{ew}^2}$
and $T_{ew}\sim 100\gev$ is the critical temperature.  The above is valid 
for $2M_W(T)\ll T\ll 2M_W(T)/\alpha_W$, i.e., it breaks down close to
$T_{ew}$.  $y$ is obtained by comparing the B violation rate to
the sphaleron rate per unit volume in eqns. (2.17) and (2.14) of
ref. \cite{amc}.  Choosing B-L=0 and three families we get y=20.
$f$ includes other factors found in eq. (2.19) of ref. \cite{amc}.

The non-perturbative lattice calculation of the sphaleron rate
per unit volume in the broken phase close to the critical temperature 
$T_{ew}$ is given by \cite{moore}
\be
\Gamma_{sph}/V=x \alpha_W^4 T^4\, ,
\ee
where $x$ varies from $10^{-8}$ to $10^{-5}$ for different values of the 
couplings in the theory.  Therefore $\Gamma(T)$ is given by
\be
\Gamma= x y \alpha_W^4  T\, .
\ee

We can model the rate of baryon number violation for the range
$T=100\gev-T_U$ as
\be
\Gamma(T)=x y \alpha_W^4 T exp[-E_{sph}/T] T_{ew}^7/T^7
\ee
where we ignore the temperature dependence in $M_W$.  This should hence give 
us an upper bound on the actual rate.

We then find that the reduction in the baryon number density due to sphaleron
processes is negligible.  (The fraction of baryons destroyed before
the temperature of the expanding plasma becomes $T_U$ is $10^{-10}$ and
$10^{-11}$ for $T_U$ equal to 1 \gev and 10 \gev. respectively.)
This is true even for much larger values of $x$, including the larger 
perturbative estimates of $x$ shown in fig. 4 of ref. \cite{moore}.

The expression for the asymmetry in the second order phase transition case, 
adapted from the results of ref. \cite{prokopec}, takes into account  
any destruction of the asymmetry when the value of the asymmetry at any  
time becomes greater than the equilibrium value.

The above arguments, coupled with the observation that our model works for
a second order electroweak phase transition, implies that
the constraint of a lower bound on the Higgs mass does not
apply to our scenario.  However, as we have pointed out earlier, 
insufficient CP violation in the Standard
Model requires one to consider extensions of the Standard Model.  

\section{Conclusion}

The baryon asymmetry that is created in our scenario is inhomogeneously
distributed in the universe.  However an estimate of the distance traversed 
by the baryons beyond $r_U$ (in the plasma motion case)
due to random Brownian motion indicates that the baryons
produced in our scenario homogenize themselves on time scales of the
order of the lifetime of the black holes.  Particles traveling through
a plasma suffer collisions and the actual distance traveled by
a particle in the time $\tau$ is $\sqrt N l_p$, where $N$ is the number of
collisions and $l_p$ is the mean free path\cite{Feynman}.  Now 
$N=\tau/(l_p/c)$.  Therefore $\sqrt N l_p= (\tau c l_p)^{1\over2}$.
The mean free path=$1/(\sigma n)$ and, as stated earlier,
$\sigma n\sim0.02T$.  If we substitute the black hole lifetime for $\tau$ then
the distance traversed by the baryons beyond $r_U$ due to random Brownian 
motion is $5\times 10^9\gev^{-1}$ and $1\times10^8\gev^{-1}$ for ambient 
temperatures of 1 \gev and 10 \gev respectively.  Since the distance between 
black holes is $5\times10^{9} \gev^{-1}$
and $1\times10^8\gev^{-1}$ for ambient temperatures of 1 \gev and 10 \gev
respectively and the inverse Hubble parameter is of the order of the black
hole lifetime this indicates that the baryon density inhomogeneities can 
be wiped out on time scales of the order of few times the black hole lifetime.

Neutrino inflation and baryon diffusion can also
homogenize any baryon number density fluctuation prior to
the onset of primordial nucleosynthesis.  Neutrino inflation is the
process of heating up of overdense baryon density lumps in the early
Universe by neutrinos which causes the
lumps to expand, or inflate, thereby reducing their overdensity.  Besides
this mechanism, neutrons and protons also diffuse out of overdense baryon
density lumps.  Mechanisms of diluting overdense baryon density lumps are
discussed in ref. \cite{JF}.  The above arguments also apply to the 
diffusion scenario where the size of the baryon density inhomogeneity is
given by $\r100$.

As we mentioned earlier, the scenario in ref. \cite{nagatani}
has similar features to our work.  
In this paper one obtains the temperature profile outside the
black hole in a manner similar to that used for stellar interiors.  As we
have pointed out, the assumption that heat transfer is primarily due to 
diffusion, as in stellar interiors, is not necessarily valid for the 
plasma surrounding a primordial black hole.

There are many open questions which remain to be explored.  The issue of
black hole formation with the required number density and mass spectrum
for our scenario requires to be discussed in the context of some realistic
model such as a first order phase transition occurring earlier in the Universe.
Also one must keep in mind observational constraints on massive black 
holes \cite{carrreview,macgibboncarr,barrowcopelid}
which may lead to strong restrictions on the spectrum of masses of
primordial black holes.  A more complete numerical calculation can give
us an estimate of the temperature of the plasma close to the black hole and
fix $\eta$.  Finally viscous effects can be included in the relativistic
Euler equation to more accurately ascertain the velocity of the plasma.

\vskip .2in
\centerline {\bf ACKNOWLEDGEMENTS}
\vskip .1in

Salient features of this paper were presented earlier by R. Rangarajan
at the workshop on `Standard Model Physics' at the Institute
for Theoretical Physics, University of California, Santa Barbara, USA 
in June '94. Also, a preliminary version of this paper was presented 
by A.M. Srivastava at the workshop on `Quark-Gluon Plasma and Phase 
Transitions in the Early Universe' at Puri, India in December '95. We thank
the participants of these workshops for many helpful remarks and 
suggestions. In particular we thank David Kaplan and Ann Nelson for many
useful suggestions. We also thank Sanatan Digal for useful discussions.  
R.R. would like to thank Jitesh Bhatt for very constructive discussions.

%\newpage

%\newpage

\vskip .3in
\centerline {\bf FIGURE CAPTIONS}
\vskip .1in

Fig.1: Effective potential when the region superheats with $<\phi> = v$
becoming metastable.

Fig.2: (a) Nucleation of $<\phi> = 0$ bubble in the metastable phase
with $<\phi> = v$. (b) Nucleation of $<\phi> = v$ bubble inside a shrinking
$<\phi> = 0$ bubble. 

\newpage

%%%%%%%%%%%%%%%%%%%%%%%%%%%%%%%%%%%%%%%%%%%%%%%%%%%%%%%%%%%%%%%%
\vskip -0.25in
\begin{figure}[h]
\begin{center}
\leavevmode
\epsfysize=15truecm \vbox{\epsfbox{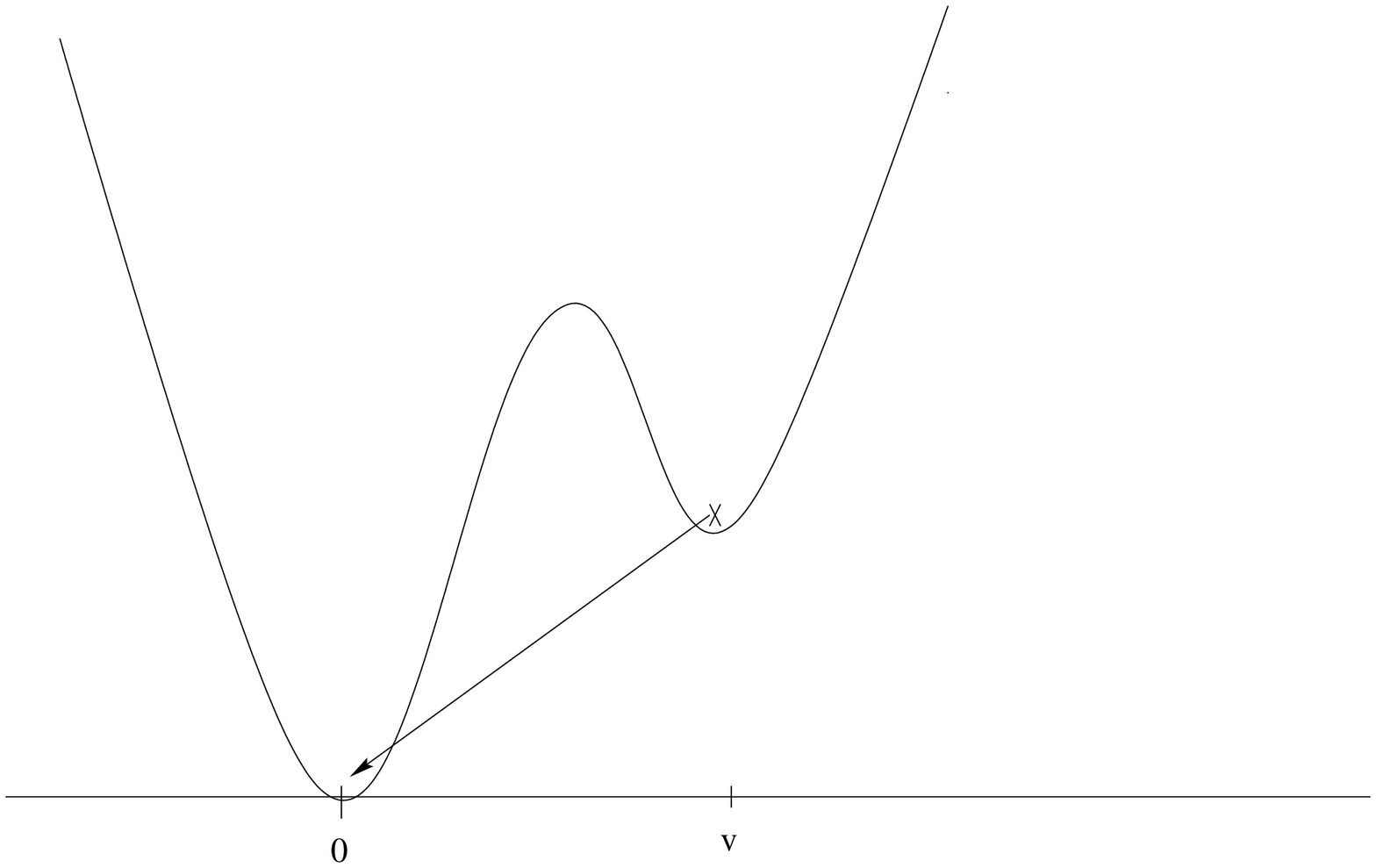}}
\end{center}
\vskip 1.0in
\caption{}
\label{Fig.1}
\end{figure}
%%%%%%%%%%%%%%%%%%%%%%%%%%%%%%%%%%%%%%%%%%%%%%%%%%%%%%%%%%%%%%%%%%

\newpage

%%%%%%%%%%%%%%%%%%%%%%%%%%%%%%%%%%%%%%%%%%%%%%%%%%%%%%%%%%%%%%%%
\vskip -0.35in
\begin{figure}[h]
\begin{center}
\leavevmode
\epsfysize=15truecm \vbox{\epsfbox{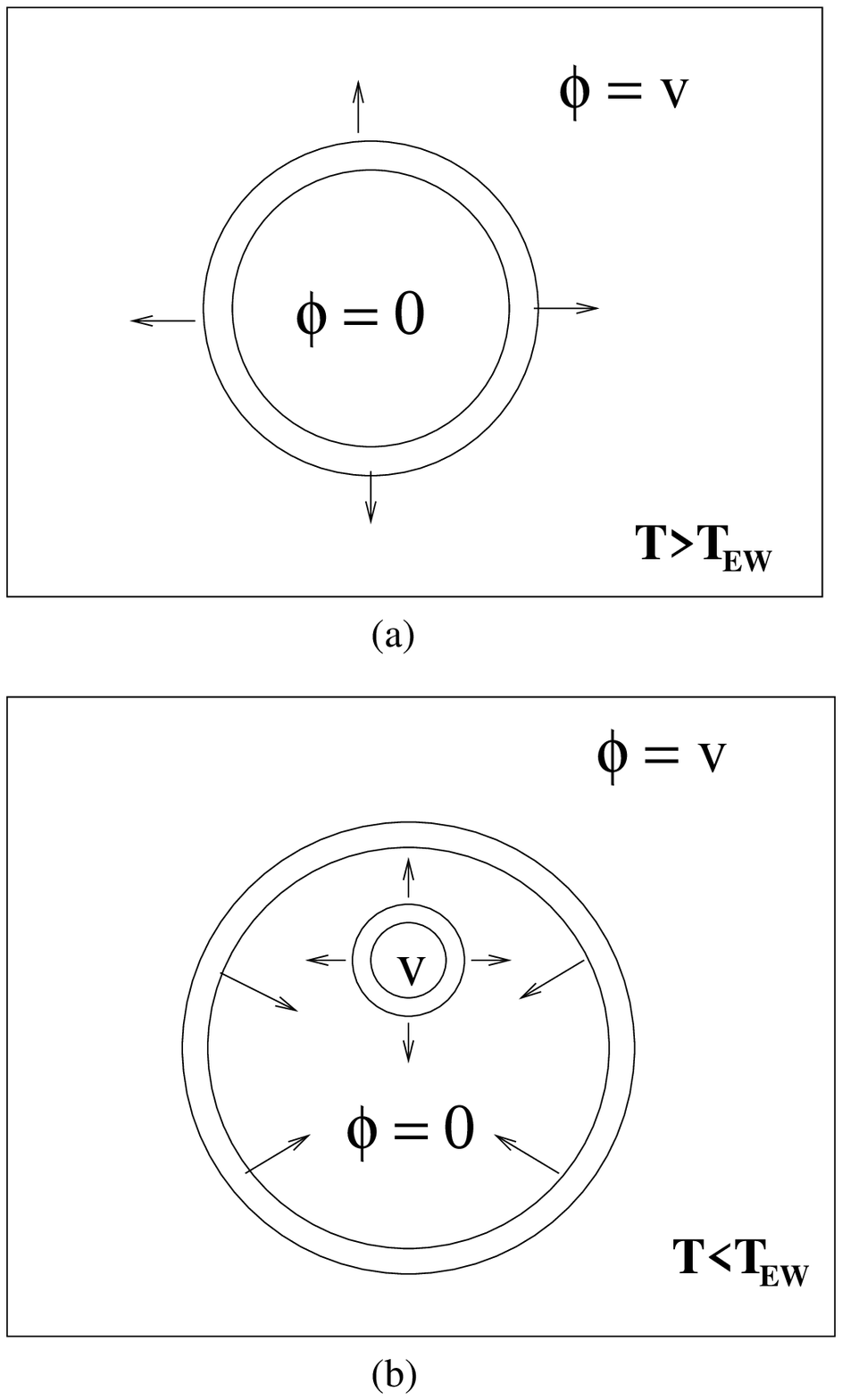}}
\end{center}
\vskip 1.0in
\caption{}
\label{Fig.2}
\end{figure}
%%%%%%%%%%%%%%%%%%%%%%%%%%%%%%%%%%%%%%%%%%%%%%%%%%%%%%%%%%%%%%%%%%

\end{document}